# Partition of coating agents between nanoparticle interfaces and polymer in nanocomposites

Dafne Musino[1], Julian Oberdisse[1], Michael Sztucki[2], Angel Alegria[3], and Anne-Caroline Genix[1*]

[1] *Laboratoire Charles Coulomb (L2C), Université de Montpellier, CNRS, F-34095 Montpellier, France*

[2] *European Synchrotron Radiation Facility, 71 avenue des Martyrs, CS 40220, 38043 Grenoble Cedex 9, France*

[3] *Departamento de Fisica de Materiales (UPV/EHU) and Materials Physics Center (CSIC-UPV/EHU), Paseo Manuel Lardizábal 5, San Sebastian 20018, Spain*

\* *Author for correspondence: anne-caroline.genix@umontpellier.fr*

**Abstract**

Industrial and model polymer nanocomposites are often formulated with coating agents to improve polymer-nanoparticle (NP) compatibility. Here the localization of silane coating agents in styrene-butadiene nanocomposite is investigated through the segmental dynamics of the polymer matrix by broadband dielectric spectroscopy (BDS), allowing the detection of silanes in the matrix through their plasticization effect. This acceleration of dynamics was followed via the shift of $\tau_{max}$ of the α-relaxation induced by the presence of coating agents of different molecular weight and quantity, for different amounts of incorporated colloidal silica NPs (R ≈ 12.5 nm, polydispersity 12%). Any noteworthy contribution of interfacial polymer layers on $\tau_{max}$ has been excluded by reference measurements with bare NPs. Our approach allowed quantifying the partition between the matrix and the NP interfaces, and was confirmed independently by calorimetry. As a control parameter, the silane grafting reaction could be activated or not, which was confirmed by the absence (resp. presence) of partitioning with the matrix. Our main result is that in the first steps of material formulation, before any grafting reaction, coating agents both cover the silica surface by adsorption and mix with the polymer matrix – in particular if the latter has chemical compatibility via its functional groups. Silane adsorption was found to be comparable to the grafted amount (1.1 nm$^{-2}$), and does not increase further, confirming that the plateau of the adsorption isotherm is reached in industrial formulations. These results are hoped to contribute to a better understanding of the surface reactions taking place during complex formulation processes of nanocomposites, namely the exact amounts at stake, e.g., in industrial mixers. Final material properties are affected both through NP-matrix compatibility and plasticization of the latter by unreacted molecules.

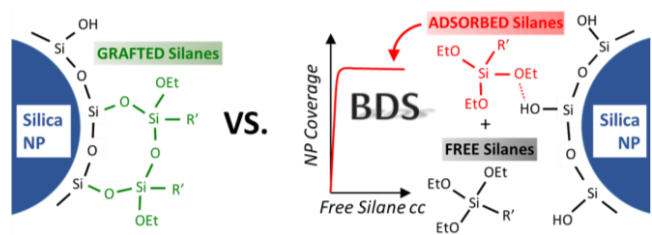

**Introduction**

Polymer nanocomposites (PNCs) are fascinating materials made of hard nanoparticles (NPs) embedded in a polymer matrix. [1-4] The performance of such materials in terms of mechanical (or electrical, permeation, etc. …) macroscopic properties depends in a crucial way on the state of dispersion of the NPs in the matrix. [5-9] The latter is controlled by processing conditions, [10] and by the chemical compatibility between the polymer and the NPs. [11-12] The fine-tuning of their interfacial properties by surface modification with small molecules as coating agents is thus a method of choice used both in fundamental studies [13] and industrial applications. [14-15] Silane molecules possessing a short aliphatic chain and graftable head groups – typically triethoxysilanes – are most often used, [16-17] in particular for hydrophobization of hydrophilic NP surfaces like metal oxides or silica. For the former, phosphonic acids can also be envisaged. [18]

In industrial processes, liquid silane coating agents are usually introduced together with NPs and the polymer in solid-state mixers. Their quantity is proportional to the amount of nanoparticles, and a typical empirical value is 8% in mass ratio between an octyl-triethoxysilane (commonly called octeo), and nanoparticles in the 10 nm range – size matters as it depends on the specific NP surface. This empirical quantity ensures successful mixing of particles with the polymer. In recent work, [16] we have studied the impact of octeo content (from 0 up to 8%w) in simplified industrial PNCs showing that compatibilization of the hydrophilic filler surface with the hydrophobic matrix induces the formation of smaller aggregates, i.e., a better dispersion.

In solid-state mixing, high temperatures are reached due to the high viscosity and strong energy input, enabling not only the grafting reaction of the silane, but also possible grafting reactions of functionalized polymer chains onto silica by condensation with the surface silanols. The success of surface modification by the silane is directly related to the interfacial concentration of silanes, which also possess good miscibility with the polymer matrix. Therefore, the question addressed here is the determination of the partition of the coating agents between the NP interface (*adsorbed* silanes) and the polymer matrix (*free* silanes), in a polymer-NP mixture *before* any grafting reaction. We have chosen to simplify the experimental system to a model system, different from industrial ones in two main aspects: first, the silica are well-defined colloidal NPs, facilitating a structural study by scattering, and secondly, mixing is performed by solvent casting. This allows removing the high-temperature stage, and thus any complication caused by additional grafting reactions. On the other hand, grafting of the coating agent can be controlled in solvent, namely by pH and temperature. In presence of water, free silanes may polycondense. Throughout this study, this is excluded due to the lack of water molecules in the hydrophobic polymer matrix.

The difficulty of assessing the silane concentrations in the bulk of a PNC, either at the nanoparticle interface or in the matrix, has impeded any direct experimental approach so far, and sample formulations remain essentially empirical. Surface modification itself may be performed in suspension. The latter may be purified to remove non-covalently grafted molecules, e.g., by dialysis, and grafted quantities determined by thermogravimetric analysis (TGA). [17] Such an approach is impossible with final PNCs, where polymer and silane degrade in the same temperature range.

The approach chosen here is based on the modification of the segmental dynamics of the polymer matrix by the presence of small molecules, usually with a plasticization effect tending to speed up dynamics, and thus reduce the glass-transition temperature, $T_g$, of the material. Broadband dielectric spectroscopy (BDS) provides access to the frequency-dependent response of molecular dipoles to an external electric field. [19] The segmental ($\alpha$-) dynamics related to the glass transition of polymers can thus be probed, as well as secondary processes like the β-relaxation characteristic of local motions, for instance side-group reorientations. These processes can be followed in frequency and temperature. The difficulty is that in absence of spatial resolution, other charge transport processes like ionic conduction or interfacial polarization stemming from incorporated NPs may overlap in the same frequency window. [20] In the present article, however, the α-relaxation is sufficiently well characterized even in presence of NPs that it can be finely described using standard fitting procedures. In our system, quantification of the silane partition is possible because the nanoparticles do not strongly influence the polymer dynamics, and it is checked that the neat matrix and the bare (i.e., no coating agent) nanocomposite have a close dynamical response in terms of the maximum of the distribution of relaxation times, $\tau_{max}$, as compared to the shifts induced by the plasticization. The contribution of any interfacial polymer layer is at most a small broadening of the shape of the relaxation function, without shifting the distribution [21] – contrarily to plasticization as shown in this article. The position of the α-relaxation in the frequency domain thus gives access to the bulk response of the nanocomposite, and thus the matrix composition in terms of presence of free (i.e., non-adsorbed) silanes.

The existence and dynamical signature of interfacial polymer layers surrounding the nanoparticles in PNCs has attracted considerable interest in the past. [22-24] In attractive NP-polymer systems like poly(2-vinylpyridine) (P2VP)/silica and poly(vinyl acetate) (PVAc)/silica, typical thicknesses range from 2 to 5 nanometers. [25] In BDS studies, their contribution usually leads to an additional and noticeable contribution to the dielectric response visible on the low-frequency side of the α–process, and a reduction in the intensity of the α–peak (beyond silica contribution). In this case, the overall segmental dynamics evaluated by means of the apparent peak maximum in the dielectric spectra is shifted towards longer time scales and a decrease in the calorimetric $T_g$ is observed. [26] In typical rubber systems, like the one studied here, however, enthalpic interactions between the polymer segment and the NP surface (hydroxyl groups) are weak and the influence of the interfacial polymer layer is extremely low, as shown by incoherent neutron spin-echo [21] or low-field NMR experiments. [27-28] In the latter, at most a few percent of immobilized polymer forming nanometer-thick shells may be identified.

This article is organized as follows. First, the dielectric response of the polymer matrix in presence of coating agents – but without the NPs to be coated – is studied. The result is a calibration curve, which relates the amount of silane molecules in the matrix to the position of the maximum of the α-process described by $\tau_{max}$. Then, polymer nanocomposites without grafting of the silanes are investigated for different silica volume fractions, at fixed silane-to-silica mass ratio. Using the calibration curve, the amount of free silane in the matrix is determined. By mass conservation the adsorption isotherm on the nanoparticles within the polymer nanocomposite can then be constructed and compared to grafted amounts if the surface modification with silanes were performed first. Finally, some theoretical considerations based on the Vogel-Fulcher-Tammann-law are discussed, showing that while both calibration curve and PNC-response shift with temperature (for a fixed partition of coating agent), they can be used at any temperature, leading to the same isotherm. These results are further confirmed by the evolution of the calorimetric $T_g$.

**Materials and Methods**

**Sample formulation.** Silica NPs are rather monodisperse colloidal particles suspended in water (Ludox TM40). Their dispersion in size has been measured by SAXS and can be described by a log-normal distribution function with parameters $R_0$ = 12.5 nm, log normal polydispersity σ = 0.12. This leads to an average surface and volume of <A> = 2020.8 nm$^2$, and <V> = 8728.7 nm$^3$. Bare NPs have been transferred in three steps into organic solvents for mixing with the polymer, first into ethanol and then twice into MEK, by dialysis (24h for each step), and sonicated for 30 min at room temperature. Success of the dialysis was evaluated by NMR, with remaining quantities of ethanol and water below 2.5%.

The polymer matrix is made of random styrene-butadiene (SB) copolymers with 19.1%w styrene units, and 80.9%w butadiene, out of which 42.6%w are 1,4-units, the rest 1,2. Two uncrosslinked matrices were studied, non-functionalized (termed SB_NF) and end-functionalized styrene-butadiene (SB_F) that bears a single silanol end-function. They are of nearly identical molecular weight, namely 177 and 170 kg/mol, respectively, and polydispersity index of 1.02. Both were synthesized by Synthos following the protocol described in ref [29]. The activation of the grafting reaction of the polymer onto the silica is usually triggered by the high temperatures (T > 150 °C) in internal mixers submitting the material to high shear. Given that such high temperatures have been avoided in the present study by following a solvent-route, no effect of the end-functionalization on filler dispersion is expected.[29] Moreover, the SiMe$_2$OH-end group (often termed D3 in the literature), was found to have no influence on the polymer segmental dynamics by BDS (see results below). However, it will be shown here that the partition of the silane coating agents does depend on the presence of D3 functional group within the polymer melt.

The silane molecules used as coating agents (CA) are octyl-triethoxysilane C$_8$ (molecular weight Mw = 276 g/mol), dodecyl-triethoxysilane C$_{12}$ (Mw = 332 g/mol) and octadecyl-triethoxysilane C$_{18}$ (Mw = 417 g/mol). C$_8$ was supplied by Sigma-Aldrich, and C$_{12}$ and C$_{18}$ by Alfa Aesar. The quantity of silane in the samples is given by the fixed ratio with respect to the mass of silica nanoparticles (8%w, 9.6%w, 12%w, for C$_8$, C$_{12}$, and C$_{18}$, respectively, ensuring isomolarity), which is kept constant for all experiments with silica NPs discussed here, unless otherwise stated. The silanes may or may not be covalently grafted on the silica. This surface modification is activated by a 24h-reaction on a silica suspension in an ethanol-water mixture (63%v ethanol), at pH 9 and T = 50°C, [17] and followed by the same phase transfers (pure ethanol and then MEK) than bare NPs. As the latter are based on dialysis, this purifies the suspension, and free silane molecules in the solvent are removed. The surface-modified silica NPs were characterized after washing cycles by centrifugation and redispersion in ethanol. Grafting densities were measured by TGA (Mettler Toledo, 5 K/min from 35 to 900 °C under air, 60 mL/min) and amount to 1.1 nm$^{-2}$ assuming on average hydrolysis of two ethoxy groups on the NP surface. We checked that the same values were obtained by characterizing the grafting density after transfer in MEK. The grafting density corresponds to a loss of ca. 0.55 nm$^{-2}$ with respect to initially introduced coating agent. If grafting is to be avoided, the silanes are added to the NP suspension after the last dialysis against MEK, corresponding to a quantity with respect to the silica surface of 1.65 nm$^{-2}$, and there is no further purification. Suspensions thus carry either bare (no silane), grafted, or non-grafted NPs. In the latter case, silanes are present in solution and/or adsorbed to the NP surface.

Polymer nanocomposites were prepared at silica volume fractions ranging from 0% to ca. 20%v. PNCs were formulated by mixing appropriate solutions of polymer with bare, non-grafted, or grafted silica suspensions in the common solvent (MEK), followed by casting on a Teflon mould, for 24h at 50°C and drying for 24h under vacuum at room temperature. The real volume fractions $\Phi_{NP}$ were determined by TGA from the weight loss at 900°C under air. After nanocomposite formulation, the amount of coating agent in the matrix may be non-zero due to the possible partition of the silane between the silica surface and the matrix polymer as discussed in this article. Assuming all coating agent to be in the matrix, the nominal matrix concentration of silane ζ is expressed by the corresponding mass fraction

$$\zeta = \frac{m_{CA}}{m_{CA}+ m_{SB}} \quad (1)$$

This nominal ζ needs to be differentiated from $\zeta_{free}$ determined in this article, which corresponds to the free silane concentration really observed within the polymer matrix in PNCs. The nominal ζ values for selected PNC samples are given in Table 1.

**Table 1:** Formulation of $C_{18}$-nanocomposite samples. Matrix type, grafting reaction, silica volume fraction from TGA, silane-silica mass ratio, and nominal concentration of silane in matrix ζ.

| Matrix | Grafted | $\Phi_{NP}$ (%v) | $C_{18}$/silica ratio (%w) | ζ (%w) |
|---|---|---|---|---|
| SB_F |  | 1.6 | 12 | 0.3 |
|  |  | 5.0 | 12 | 1.6 |
|  |  | 7.5 | 12 | 2.5 |
|  |  | 10.3 | 12 | 3.4 |
|  |  | 11.8 | 12 | 4.3 |
|  |  | 14.5 | 12 | 5.4 |
|  |  | 11.3 | 0 | 0 |
|  |  | 20.4 | 0 | 0 |
|  |  | 9.6 | 18 | 5.3 |
|  |  | 9.9 | 24 | 7.2 |
|  | yes | 11.2 | 12 | 3.4 |
|  | yes | 20.7 | 12 | 7.6 |
| SB_NF |  | 10.7 | 12 | 3.4 |
|  |  | 10.5 | 18 | 5.3 |
|  |  | 10.1 | 24 | 7.2 |
|  | yes | 9.9 | 12 | 3.4 |

**Broadband dielectric spectroscopy.** The real and imaginary parts of the dielectric permittivity, $\varepsilon^*(\omega) = \varepsilon'(\omega) - i\varepsilon''(\omega)$, were measured using a Novocontrol Alpha spectrometer between T = 153 to 333 K, in the frequency range from f = $10^{-2}$ to $10^6$ Hz ($\omega = 2\pi f$), after a short equilibration at 333 K in the BDS cryostat under nitrogen atmosphere. Samples placed between two plate electrodes were typically 0.2 mm thick. Temperature control was ensured by a Novocontrol Quatro cryosystem leading to a precision better than 0.1 K. In practice, the α-relaxation was well positioned in the frequency window between T = 253 and 278 K. The imaginary and real parts of the matrix spectra were fitted by a Havriliak-Negami (HN) function with a single process describing the α-relaxation of dielectric strength Δε and characteristic time $\tau_{HN}$, and a purely dissipative d.c. conductivity term leading to a 1/ω dependence [19]

$$\varepsilon^*(\omega) = \varepsilon_\infty + \frac{\Delta\varepsilon}{[1+(i\omega\tau_{HN})^\gamma]^\delta} - i\frac{\sigma_{dc}}{\varepsilon_0\omega} \quad (2)$$

γ and δ are the width and asymmetry parameters of the HN distribution, respectively. For nanocomposites, a second HN function was added to describe the Maxwell-Wagner-Sillars (MWS) process associated with polarization effects in the presence of filler.[30] In our previous studies, such MWS processes have been described in detail following the same approach. For the purpose of well-describing the α-relaxation position $\tau_{max}$ pursued in the present paper, it is sufficient to describe the high-frequency tail of the MWS process, keeping fixed δ = 0.58 for its asymmetric broadening as obtained from higher T measurements. In all cases, fitting was carried out simultaneously on ε'(ω) and ε''(ω). The segmental relaxation reported in this paper is defined by $\tau_{HN}$ related to the peak position of the well-defined process in frequency $f_{max}$, which is used to determine the relaxation time $\tau_{max} = 1/(2\pi f_{max})$. The temperature dependence of $\tau_{max}$ is usually plotted as a function of 1/T to evidence non-Arrhenius behavior. The functional dependence is given by the phenomenological VFT equation [19]

$$\tau_{max} = \tau_0 \exp\left(\frac{DT_0}{T-T_0}\right) \quad (3)$$

where $\tau_0$ is the relaxation time in the high-temperature limit, $T_0$ is the position of the divergence, and D parametrizes the deviation from Arrhenius behavior related to the glass fragility. [31-32] The VFT equation can be used to determine the dielectric glass-transition temperature as the temperature corresponding to $\tau_{max}$ = 100 s.

**Differential scanning calorimetry (DSC).** Temperature-modulated measurements (TM-DSC, Q2000 TMDSC, TA Instruments, heating rate 3 K/min, modulation amplitude ±0.5 K, period 60 s) were performed in order to follow the calorimetric $T_g$, and thus evidence the plasticization effect. $T_g$ was defined as the inflexion point temperature in the reversible heat flow upon heating.

**Small-angle X-Ray Scattering (SAXS).** Experiments were performed on beamline ID02 [33] at ESRF (Grenoble, France). The wavelength was 1.0 Å (12.46 keV), and three sample-to-detector distances were used (SDD = 1 m, 10 m, 30 m). The nominal q-range was $2 \times 10^{-4}$ to 0.5 Å$^{-1}$, but due to strong low-q scattering from matrix heterogeneities, $q_{min}$ was set to $10^{-3}$ Å$^{-1}$. Background corrections and calibration to absolute intensity were performed following standard ESRF procedures.

**Results**

With the two types of matrices, end-functionalized (SB_F) or not (SB_NF), and the coating agents with different alkyl lengths ($C_8$, $C_{12}$, and $C_{18}$), many different samples can be formulated, at different silica contents, as given in Table 1. Here we first focus on the silica-free matrices, showing that end-functionalization does not influence the α-relaxation of the polymer, whereas the silanes – exemplarily discussed for $C_{18}$ but others are also included in the main result – speed up matrix segmental dynamics.

BDS frequency scans of SB_F matrices have been performed with steps of 5 K, and the main α-process was captured best at T = 258 K. The dielectric loss $\varepsilon''(\omega)$ at this temperature is plotted in Figure 1a, for increasing matrix concentrations ζ of $C_{18}$-silane. A single relaxation peak is well-identified, which clearly shifts to higher frequencies, thus smaller relaxation times, with increasing amount of $C_{18}$. Its amplitude slightly decreases with ζ, whereas the shape becomes a little more asymmetric (decrease of δ from ca. 0.5 to 0.4 in eq 2). Within error bars, this results in a constant peak area ($\Delta\varepsilon = 0.10 \pm 0.01$) and essentially comparable shape of the relaxation time distributions. Such features reveal that both components – SB and silane – are dielectrically active and display similar relaxation times. Such a behavior was observed in other polymer/plasticizer mixtures with strong interactions,[34] as opposed to non-interacting blends characterized by distinct segmental dynamics.[35-37]

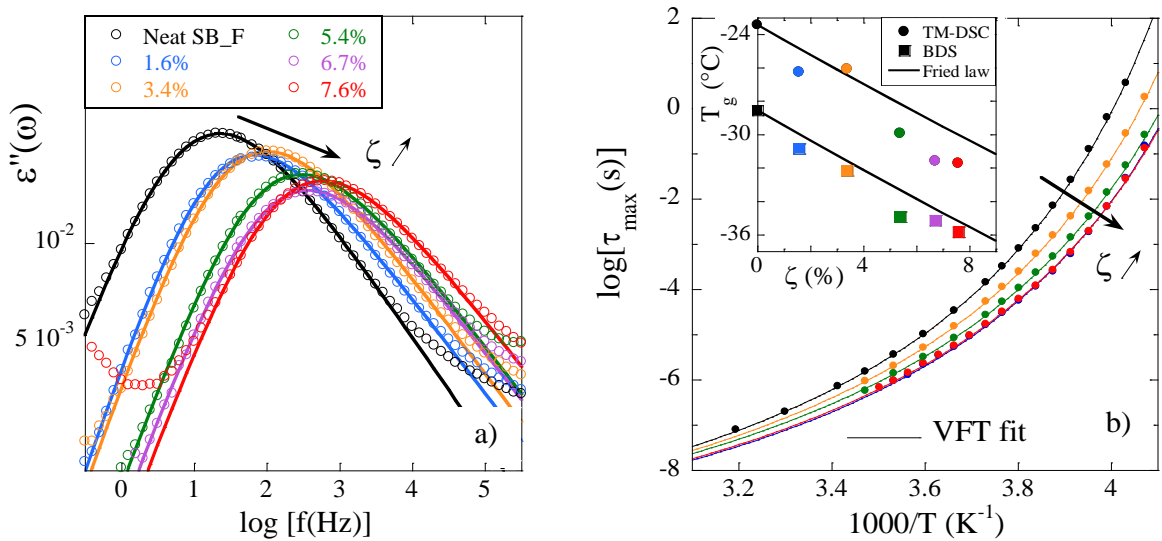

**Figure 1: (a)** Dielectric loss $\varepsilon''(\omega)$ as a function of frequency at T = 258 K of SB_F mixed with different amounts of $C_{18}$-silane at matrix concentration $\zeta$. Solid lines represent the HN fit functions. **(b)** Relaxation map for SB_F + $C_{18}$ at selected concentrations, and compared to SB_NF + $C_{18}$ at the highest $\zeta$ (dark blue symbols superimposed to the red ones for the equivalent SB_F matrix). Solid lines represent VFT fits. Inset: Calorimetric $T_g$ (circles) as a function of $\zeta$, and comparison to dielectric $T_g$ (squares). Solid lines are Fried laws [38] (the measured calorimetric $T_g$ of $C_{18}$-silane is 190 K).

The acceleration of the dynamics in Figure 1a is typical for a plasticization effect, which can also be seen by the shift in $T_g$ observed by temperature-modulated DSC. In the inset of Figure 1b, the calorimetric $T_g$ is shown to decrease by up to 10 K with added silane, i.e., increasing the silane CA concentration $\zeta$. In Figure 1b the relaxation maps – the characteristic time $\tau_{max}$ vs. 1/T – of the pure SB_F matrix and matrices with added $C_{18}$-silanes are compared. In this representation, the acceleration by the silane is seen to operate at all temperatures, but it is more effective at the lowest temperatures. At high T, virtually the same $\tau_0$ is reached, meaning that there is no additional effect of the small CA molecules once high mobility is reached. Also, the two polymer molecules, end-functionalized or not, have $\tau_{max}$ exactly superimposed in Figure 1b. Pure matrix dynamics are thus unaffected by the D3 functional end-group.

The characteristic times in Figure 1b have been fitted with eq 3. $\tau_0$ values were found to be scattered around the value of neat SB_F, $\tau_0$ = 12 ps, and they have been fixed to this value for all samples. We have also evaluated [39] the fragility parameter, m, corresponding to the steepness at $T_g$ in the relaxation map. Increasing the plasticizer content reduces the bulk polymer fragility marginally, from 112 to 108, in agreement with model computations.[40] Such a behavior may point towards a slightly slower intermolecular cooperativity of the segmental relaxation in link with lower constraints imposed by local structure.[41] Finally, the resulting BDS-$T_g$ is found to agree well with the DSC data given in the inset, giving further credit to the effect of the coating agent on the matrix segmental relaxation.

To gain further insight into the plasticization effect caused by the CA, we focus from now on a single temperature for the different matrix concentrations of $C_{18}$, ζ. The resulting $τ_{max}$ at 258 K are plotted in Figure 2. The dynamics of the matrix as a function of ζ (plain squares) is a linearly decreasing function in this logarithmic representation, and thus indicates an exponential acceleration of $τ_{max}$ proportional to exp(-$z_m$ζ), where $z_m$ is some constant characteristic of the matrix-silane system. The data points are scattered around the linear law, but the speeding up by about two orders of magnitude is beyond doubt. We have checked that exactly the same relative scattering of points is obtained at all other temperatures under study, which implies that it is due to small variations in sample composition.

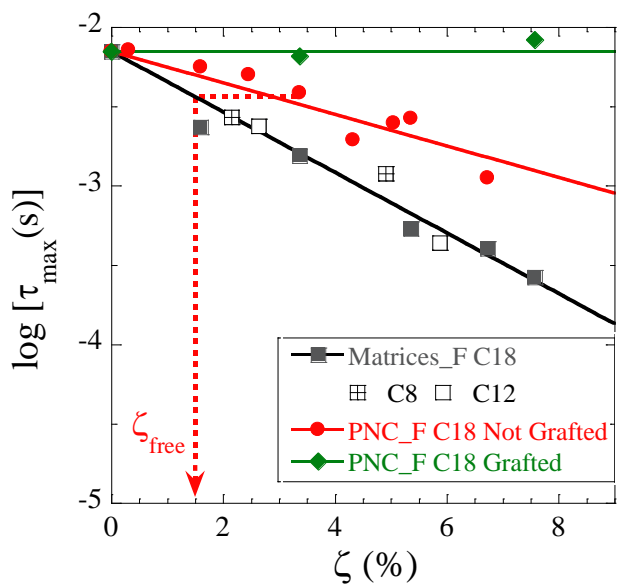

**Figure 2:** Segmental relaxation time at T = 258 K as a function of the nominal silane matrix concentration ζ in SB_F (see Table 1 for sample composition). Calibration curve for polymer with various amounts of $C_8$ (crossed empty squares), $C_{12}$ (empty squares) and $C_{18}$ (plain squares) in SB_F matrix. The solid black line represents a linear fit to all silanes in log-representation ∝ exp(-$z_m$ζ). The data for PNCs without (resp. with) activation of the grafting reaction are superimposed (red circles, resp. green diamonds) for nominal ζ in samples, together with a linear fit $τ_{max}$ ∝ exp(-zζ) for the non-grafted PNCs. The construction (red dotted lines) illustrates the determination of the effective $ζ_{free}$ within the matrix in presence of NPs.

In Figure 2, additional data for two other silanes, $C_8$ and $C_{12}$, have been included and indicated by different symbols, with the common x-axis corresponding to ζ, i.e., a mass fraction. All points fall on the linear fit, indicating that the plasticization effect does not depend on the molar quantity, but on the total mass of CA molecules. Indeed, several isomolar points for $C_8$, $C_{12}$, and $C_{18}$ are present in Figure 2, and they align corresponding to the increase in mass – which could not be the case if they had the same x-coordinate in molar units. This means that we can focus our efforts on a single coating agent, as only their mass counts. The rest of this article thus concentrates on $C_{18}$-silanes and how plasticization is impacted in PNCs.

We have shown recently that the position of the maximum of the time distribution associated with the α-relaxation is only marginally sensitive to the presence of the nanoparticles in this system. [21] There, we used a stretched exponential or Kohlrausch–William–Watts (KWW) function to describe the segmental dynamics in time domain. Applying the same approach, which amounts to coupling the shape parameters γ and δ in eq 2 (see details in SI), we can additionally compare the time distribution broadening by means of the KWW stretching exponent, $\beta_w^{1.23} = \gamma\,\delta(\gamma)$. At 258 K, $\beta_w$ is found to decrease slightly from 0.38 to 0.32 for the pure SB matrix and a PNC filled with 10%v of bare silica NPs, and it does not further evolve when increasing the silica fraction up to 20%v. The corresponding fits are given in SI with $\tau_{max}$ values and a comment on the dielectric strength. In parallel, the variation of the calorimetric $T_g$, if any, is marginal going from -23.4°C for pure SB to -24.1°C for the 20%-PNC, with ± 0.3 °C error bars. This implies a negligible impact of the interfacial polymer dynamics on the time distribution. It follows that the linear function plotted in Figure 2 using solely the maximum of the distribution at 258 K can be used as a calibration curve giving access to the amount of silane molecularly dispersed within the matrix, independently of the presence of silica NPs. Note that the T-dependence of the calibration measurements will be addressed below. In the next part, the segmental dynamics of the matrix polymer in PNCs has been evaluated.

In PNCs, due to the presence of MWS processes in the low-frequency regime, high temperatures have to be investigated first. Describing the MWS contribution allows isolating the α-process of relevance for the present study, the position of which is well defined in the frequency window at 258 K for all PNCs. In Figure 3, the dielectric response of a PNC at 14.5% silica volume fraction is shown at T = 273 K, for the case of SB_F in presence of the $C_{18}$-silane at fixed silane-to-silica mass ratio (12%), without having activated the grafting reaction. Non-grafted silanes allow studying the silane partition between matrix and NPs in PNCs. Beyond d.c. conductivity, the low-frequency upturn in ε"(ω) is attributed to a MWS process of interfacial dynamics along the nanoparticle-polymer interface, which has been related to the hydration layer around silica NPS. [28, 42-43] This polarization process can be described by adding a second Havriliak-Negami function in eq 2. The results of the fit for both ε'(ω) and ε"(ω) are displayed in Figure 3 together with the individual contributions.

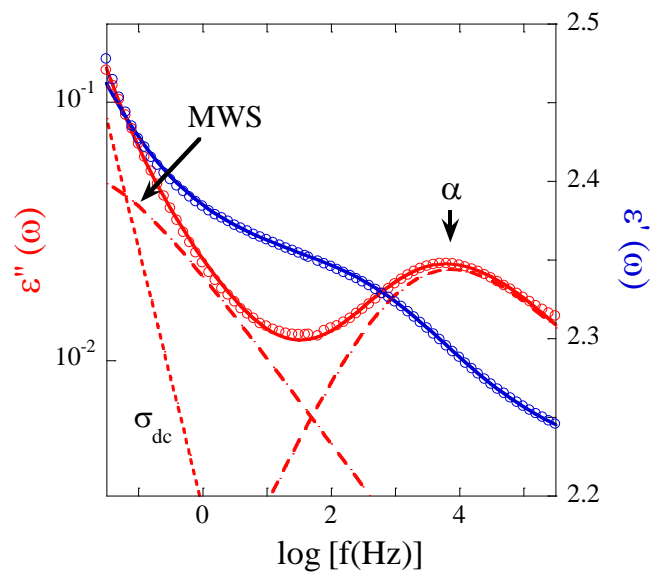

**Figure 3:** Frequency dependence of real and imaginary parts of the dielectric permittivity in PNC (T = 273 K, $\Phi_{NP}$ = 14.5%v, SB_F, non-grafted $C_{18}$). The solid lines represent the simultaneous fits of ε'(ω) and ε"(ω) based on eq 2 using two HN processes (dashed lines for MWS and segmental relaxation as indicated) and a d.c. conductivity term (dotted line).

As a result of the fits in Figure 3, we can determine the maximum of the distribution function of PNCs. A volume fraction series of nanocomposites with SB_F and non-grafted $C_{18}$-coating agent has been formulated according to Table 1, and the characteristic times obtained by BDS are shown in Figure 4 for selected compositions, including the PNC sample of Figure 3. The characteristic times are found to become smaller (i.e., faster dynamics) with increasing ζ evidencing plasticization of the polymer matrix also in presence of NPs. Their T-dependence is described using the VFT equation, giving access to the high-T limit of the relaxation time and the fragility. Within error bars, both parameters are found to be unaffected by the presence of silica, as expected from previous BDS measurements on the same system loaded with bare NPs. [21] Finally, one can also determine by extrapolation the dielectric $T_g$ in PNCs. $T_g$-values decrease by ca. 2.5 K whereas a stronger reduction of 6.4 K was measured over the same ζ range in the silica-free matrices in Figure 1b. This is a first evidence that not all silane molecules are dispersed in the bulk: a fraction of them is adsorbed on NP surfaces and does not plasticize the matrix.

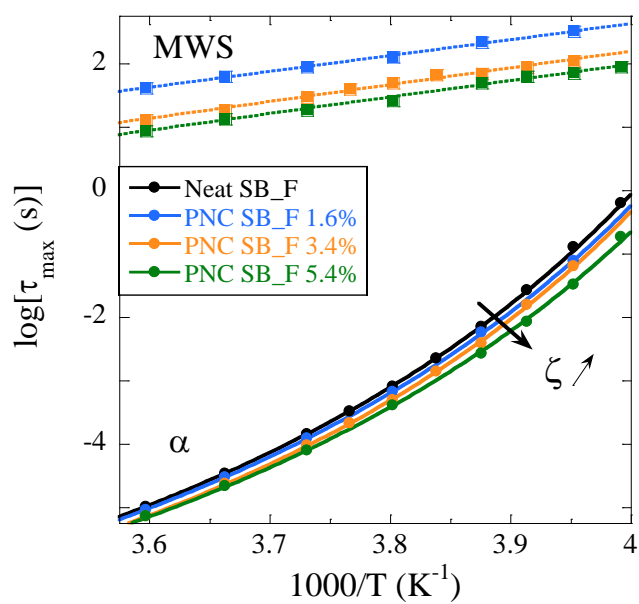

**Figure 4:** Relaxation map for nanocomposites with SB_F and non-grafted $C_{18}$ at selected silica contents ($\Phi_{NP}$ = 5.0, 10.3 and 14.5%v) corresponding to different silane concentrations ζ as indicated in the legend. It illustrates the temperature dependence of MWS (squares) and α-processes (circles). The latter is compared to neat SB_F. Solid and dotted lines represent VFT and Arrhenius fits, respectively.

Unlike the α-process, the temperature dependence of the MWS relaxation times is well described by an Arrhenius equation, $\tau_{max} = \tau_0 \exp(E_a/(k_b T))$, $k_b$ being the Boltzmann constant. The activation energy was calculated and we obtained a constant value, $E_a = (50 \pm 2)$ kJ/mol, independently of ζ and the silica content but with a fixed silane-to-silica ratio in the series (functionalized chains, non-grafted $C_{18}$). Such a value is in good agreement with values reported for simplified industrial nanocomposites based on similar SB, [28] and suggests that the chemical interphase covering the nanoparticles is not significantly modified.

To estimate the partitioning of the coating agent, the log($\tau_{max}$) values of the PNCs at 258 K from Figure 4 have been plotted onto the calibration curve of the matrix data at the same temperature in Figure 2 (red circles). The nominal ζ-value corresponds to the one calculated using eq 1. If all the coating agent went into the polymer, the same segmental relaxation time as with the silica-free matrix should be found (black line). If on the contrary none of the silane mixed with the matrix, then $\tau_{max}$ should result constant, without any ζ-dependence, as illustrated by the green line. Obviously the relaxation times of the functionalized polymer within the PNCs plotted in Figure 2 do not follow the acceleration of the dynamics of the pure matrix with added silanes, and they do not stay at constant $\tau_{max}$ either: the dynamics is accelerated, albeit considerably less – by ca. one order of magnitude – than the silica-free matrix polymer. As a result, the silane coating agents must be partitioned between the matrix and the NP interface – the latter being where they were intended to be. Our results thus provide an experimental determination of where the silane molecules are located in a nanocomposite system. This could be useful, e.g., in one-pot mixtures before grafting reactions are activated, which may be the starting point of any industrial process. For comparison, we have also added selected samples at 10%v and 20%v in silica, for functionalized polymer with grafting on the same Figure 2 (green diamonds). The segmental dynamics quantified by $\tau_{max}$ is found to be scattered around the horizontal green line, the horizontality indicating absence of partitioning with the matrix. This confirms that once grafted, the silane molecules stay at the NP interface, as expected.

Interestingly, the same behavior as observed in Figure 2 by means of the evolution of the characteristic times at a given temperature by BDS is also obtained when plotting the calorimetric $T_g$ (fixed relaxation time) versus ζ for the same samples. This equivalent calibration curve obtained independently by DSC is shown in the SI and the question of time-temperature superposition to link both calibration curves will be addressed below. It gives further support to the proposed method.

In order to determine a possible impact of chain functionalization, four combinations of matrix (SB_F or SB_NF) with activation or not of silane grafting have been studied for PNCs with a fixed silica fraction of ca. 10%v. The BDS-spectra corresponding to all four cases are reported in Figure 5a, together with the matrix response. They show that the α-relaxation appears in the same frequency window, and its position $\tau_{max}$ can be conveniently followed with varying formulation parameters. Both the MWS tail and the segmental dynamics are well captured by the model fit, and characteristic times have been reported in Figure 5b. Clearly, only the red curve presents a shift in the position of the α-process, caused by the non-grafted silanes which plasticize the SB_F matrix. Thus, only this particular sample among the four was found to present partitioning between the matrix and the NP interface.

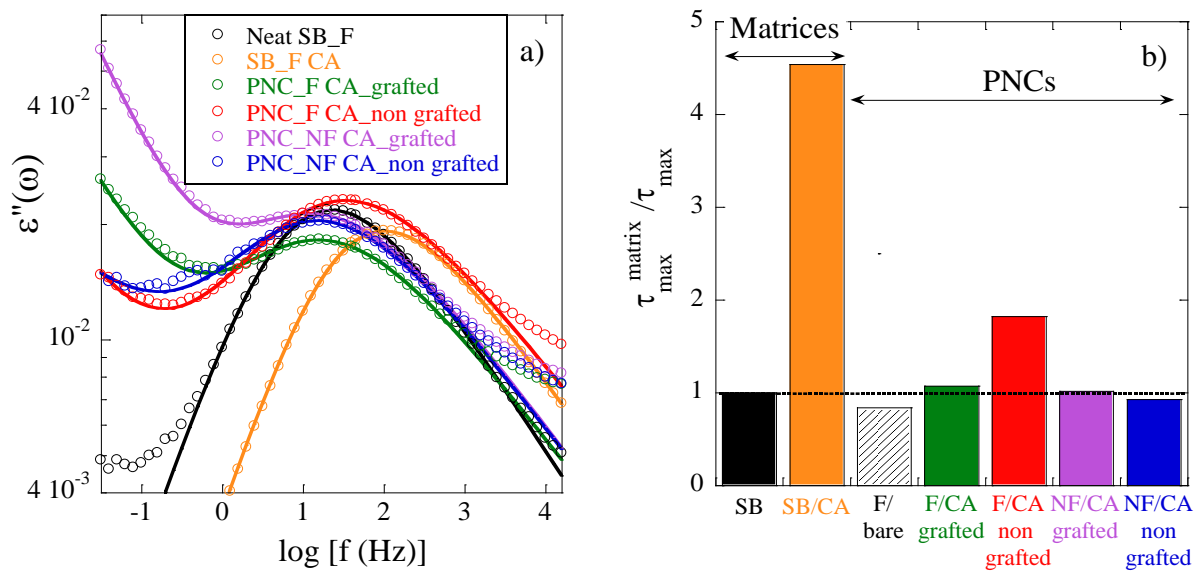

**Figure 5:** **(a)** Dielectric loss $\varepsilon''(\omega)$ as a function of frequency at T = 258 K of SB_F matrices (without and with $C_{18}$ at ζ = 3.4%) and PNCs ($\Phi_{NP}$ ≈ 10%v, SB_F or NF, same amount of $C_{18}$). Possible silane grafting is indicated in the legend. Solid lines represent the global fit based on eq 2. **(b)** Inverse of segmental relaxation times $\tau_{max}$ of the same samples, normalized to the one of pure SB_F. The result for the bare NPs shows the negligible contribution of any interfacial polymer layer.

In Figure 5b, the relative change in relaxation time with respect to the pure matrix is highlighted, represented as the inverse, such that higher values correspond to faster segmental dynamics. Note that there is no difference in dynamics between the two versions of the pure matrix, functionalized or not (Figure 1b). The strong plasticization effect of the $C_{18}$-silane in SB corresponds to an acceleration by a factor of more than four, whereas all PNCs but one show unchanged α-time scales. This emphasizes the propensity of the silanes to cover the NP interface. Only in the functionalized matrix, the non-grafted coating agents dissolve partially in the bulk, and thus accelerate its dynamics, as one can see in Figure 5b by almost a factor of two. It is also clear from Figure 2 (together with Table 1) that this factor increases with increasing silica content, meaning that more coating agent accumulates in the matrix, presumably due to the overall increasing silane concentration which is added at fixed mass ratio with respect to the NPs. Finally, in absence of coating agent ("bare" sample in Figure 5b), the $\tau_{max}$ of the segmental dynamics is found to change only marginally (see also [21]), confirming the much stronger role of the silanes in the matrix with respect to any possible impact of the silica content on $\tau_{max}$, as discussed previously. In parallel, the shape of the time distribution was hardly changed with respect to pure SB ($\beta_w$ = 0.32, resp. 0.38). This reference measurement on bare NPs shows that the effect of silica on SB-polymer dynamics is small, contrary to strongly attractive systems of the literature discussed in the introduction. If one goes into more detail also with the PNC samples with $C_{18}$-silanes in Figure 5b, fitting of the dielectric response using a KWW distribution shows that the weak broadening with silica does not further depend on the quantity of plasticizer (same stretching exponent of $\beta_w$ = 0.32), whereas the maximum of the distribution is basically independent of data analysis (see SI), i.e., with or without shape-parameter coupling.

In summary, there is no sign of any strong interfacial polymer effect on the position of $\tau_{max}$, which is thus a robust indicator: the strong shift observed in the α-relaxation in Figure 5b for samples of the type [SB_F/CA non grafted] can be attributed to the partition of the coating agent. The remaining question to be answered using the construction to determine $\zeta_{free}$ in Figure 2 is how much of the coating agent is actually present in the matrix, and how much on the silica interface. It will be addressed below after the structural analysis.

In order to exclude any major reorganization of the nanocomposites, which might be responsible of the differences observed in Figure 5, we have studied the dispersion of the NPs in the matrix. The structure of the four types of samples, i.e., grafted and non-grafted model PNCs, formulated with SB_F or SB_NF, and $C_{18}$ as coating agent has been investigated by SAXS. Due to the particle-polymer contrast and the low contributions of the interface, SAXS intensities give information on the spatial correlation between particles, and on the shape of the latter described by the form factor P(q).

The SAXS intensities normalized to the silica volume fraction of ca. 10%v are shown in Figure 6a. The scattering curves overlap rather well in the intermediate and high-q regimes, and show (small) differences at low q, indicating a slightly different state of average aggregation. The type of the matrix seems to be correlated with the upturn at very low q, the non-functionalized polymer favoring the higher increase. This corresponds to the better dispersion as observed in industrial nanocomposites,[29] but remains puzzling as no high-temperature treatment favoring polymer grafting has been performed, and will not be further discussed here. For comparison, the NP form factor P(q) has been measured independently in dilute suspension, and shifted vertically to be superimposed to the data at higher q. This allows highlighting the main structural features of the PNC samples: they are all slightly aggregated – the low-q limit exceeds the one of P(q) –, and seem to be divided in two groups. The non-grafted samples have the lower aggregation, and possibly a shallower correlation hole at q approaching the close-contact value between spheres (see arrow in Figure 6a).

In order to make aggregation directly comparable, the apparent structure factor S(q) obtained by dividing the intensities by the form factor has been plotted in Figure 6b. The two PNCs with grafted NPs are seen to have a higher nearest neighbor peak at q = 0.028 Å$^{-1}$, and a deeper correlation hole just before. This indicates statistically more closest neighbors, and corresponds to the higher average aggregation number found in the low-q range of S(q). This number is quite low for the four samples, between 2 and 6, and corresponds to a wide distribution of (small) aggregates, presumably with many isolated beads, as one may judge from the quantitative analysis of the depth of the correlation hole:

[44] the local particle volume fractions are all close to the nominal silica fraction of 10%v, indicating globally good dispersion. Consistently with the visual inspection of the structure factor in Figure 6b, a slightly higher local density is found for the two more aggregated samples, i.e., with activated grafting of the coating agents.

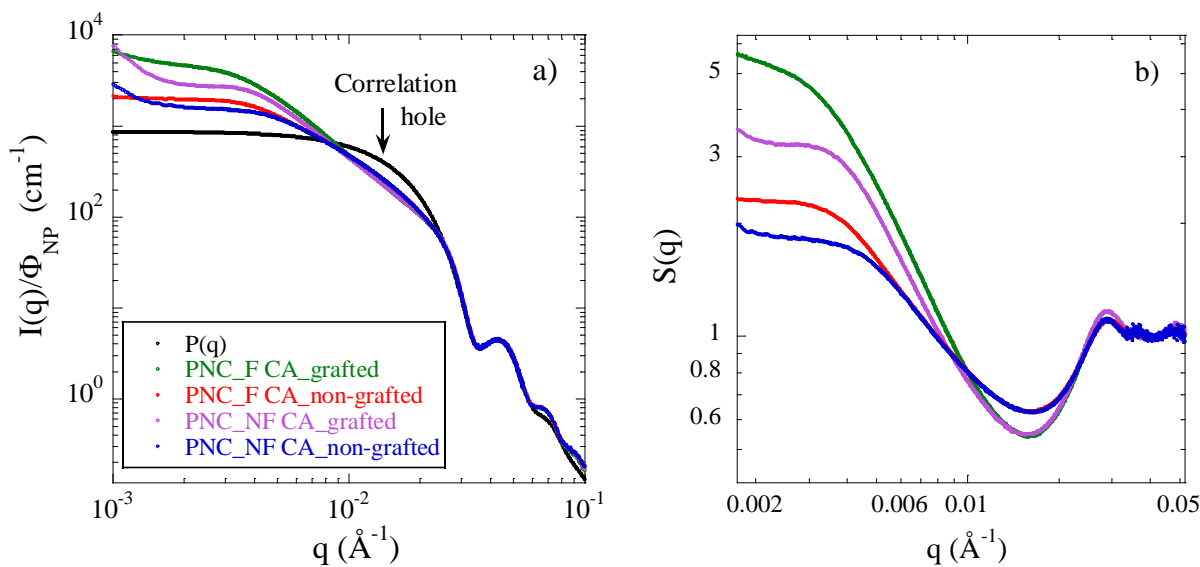

**Figure 6: (a)** Reduced SAXS intensities $I(q)/\Phi_{NP}$ vs. wave vector q for PNCs formulated with silica ($\Phi_{NP} \approx 10\%v$) grafted or non-grafted with $C_{18}$ ($\zeta = 3.4\%$) in SB_F or SB_NF, as indicated in the legend. The experimentally measured average form factor P(q) is superimposed at high q. **(b)** Apparent structure factor for the same samples.

The structural analysis performed by SAXS allows us to conclude on several points: first, NPs are globally well dispersed in the matrix, as no strong intensity increase is observed at the relevant length scales – note that $q_{min}$ corresponds to sizes of some 20 NPs in one dimension, i.e., strong aggregation would show up in our plots by increases by several decades in intensity. A modification of the dynamics caused by major structural changes of these four samples may thus be excluded. Secondly, scattering confirms the efficiency of the grafting reaction, as it has consequences on the structure. Activation of grafting induces some slight aggregation of the order of less than 10 NPs, presumably due to the formation of polycondensed patches on silica surface favoring attractive interactions as conjectured in a previous study. [17] The linear dimension of these aggregates as measured by the low-q shoulder around 0.003 Å$^{-1}$ corresponds to about only two particles as found by a simple Guinier fit, which is consistent with the low aggregation. When grafting is activated with T and appropriate pH, neighboring molecules may polycondense at the interface forming attractive patches, whereas they are simply adsorbed in absence of grafting. In case the matrix is chemically compatible with the presence of silanol end-functions, the coating agents can thus partition between the interface and the polymer, as demonstrated in Figure 5.

As a last measurement, the degree of saturation of the NP interface by the coating agents has been tested by adding higher silane-silica mass ratios (see Table 1), without activating silane grafting. The dielectric data for 10%-silica nanocomposites with a functionalized matrix are shown in Figure 7a evidencing the α-process for different ratios starting from 0, i.e., with the bare NPs. In Figure 7b, the logarithm of $\tau_{max}$ is plotted as a function of the silane-silica ratio in the same PNCs and compared to 10%-silica PNCs with a SB_NF matrix. At the silane-silica ratio of 12% – the value used throughout the rest of this article for $C_{18}$ – only the functionalized matrix partitions the silane, inducing a decrease of $\tau_{max}$ as seen in Figure 5. At higher silane-silica ratios (18%, resp. 24%) displayed in Figure 7b, the $\tau_{max}$ of both polymers decreases, and gets closer. This indicates that the additional silane is free to go into the matrix leading to stronger plasticization, even with non-functionalized polymer, presumably because the NP interface is already saturated as discussed below.

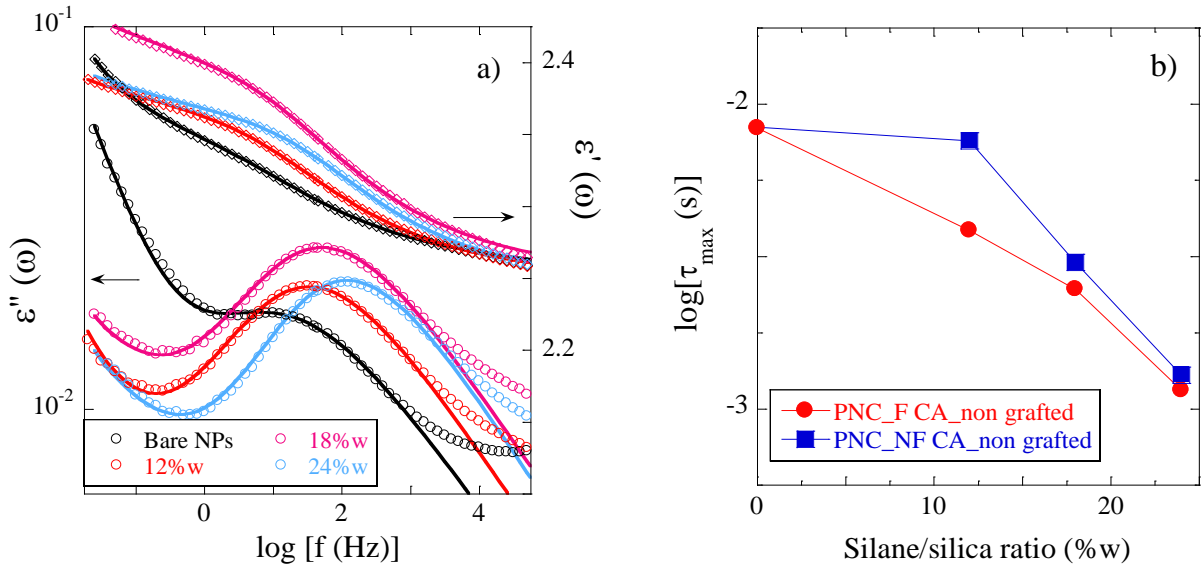

**Figure 7: (a)** Frequency dependence of real and imaginary parts of the dielectric permittivity in PNC (T = 258 K, $\Phi_{NP} \approx 10\%v$, SB_F, non-grafted $C_{18}$) for different silane-to-silica ratio as indicated in the legend. The solid lines represent the simultaneous fits of ε'(ω) and ε''(ω) based on eq 2. **b)** Evolution of the characteristic time of the α-relaxation at 258 K vs. the silane-to-silica mass ratio for the same PNC samples (circles) and corresponding PNCs in SB_NF (squares). Lines are guide to the eye.

Finally, the calibration curve in Figure 2 can be used to quantitatively determine the degree of partitioning and thus the interfacial coverage. The procedure is illustrated by the red construction (dotted lines): the comparison between the PNC relaxation time log($\tau_{max}$) and the calibration curve ($a_0 - z_m\zeta$) made with the matrix allows determining the concentration of free silanes dispersed in the bulk polymer, $\zeta_{free}$, for each nanocomposite sample

$$\zeta_{free} = \frac{a_0 - \log(\tau_{max})}{z_m} \quad (4)$$

where $a_0$ is the value of the logarithm of the relaxation time of the pure (silane-free) matrix.

Before proceeding with this point-by-point evaluation, the average evolution of the nanocomposite relaxation times described by the straight line in Figure 2 can also be used to estimate the average partition, and thus the average coverage of the NPs by the silane. Indeed, the dependence of log($\tau_{max}$) on $\zeta$ for PNCs can be described by a linear function $a_0 - z\zeta$ as shown by the red line in Figure 2, where z now describes the average evolution of the dynamics in the nanocomposite. The dotted construction in Figure 2 being based on two linear functions of negative slope z (resp. $z_m$), it immediately follows that the concentration of silane freely dispersed in the matrix is a constant fraction of the nominal concentration $\zeta$, and is given by

$$\zeta_{free} = \frac{z}{z_m}\zeta \quad (5)$$

From z = 9.92 and $z_m$ = 19.08 ($\tau_{max}$ in seconds), it follows that the real matrix concentration of silanes $\zeta_{free}$ is 52% of $\zeta$ in the functionalized matrix. One can thus conclude that the typical coverage of the silica NPs by adsorbed silanes is of the order of 50% of the nominally introduced silanes in SB_F without grafting, regardless of the silica content.

We now turn to the influence of temperature on these results. Up to here we have only analyzed data at fixed T. As the temperature-dependence of both calibration matrix and nanocomposites are known (see Figures 1b and 4, respectively), the functional dependence of the corresponding VFT-parameters log $\tau_0$, $D_0$ and $T_0$ (eq 3) on $\zeta$ can be derived, provided that the sample composition does not evolve with T. Details are given in the SI. It is found that it is sufficient to describe $D_0$ by a constant as its variation is weak, and $T_0$ by a linearly decreasing function $T_0' - a\zeta$, where $D_0$ and $T_0'$ are identical for all samples by construction. Log $\tau_0$ was found to be constant. It results that the dependence of the logarithm of $\tau_{max}$ factorizes into a T-dependent term, and a linear term in $\zeta$ of the form $a_0 - z\zeta$, as found experimentally in Figure 2. The prefactor and slopes thus evolve with T via the general prefactor, as also shown in the SI, but the construction to determine $\zeta_{free}$ remains valid at any temperature, and leads to the same result. Moreover, this theoretical description based on the VFT-parameters describes well the experimental calibration curve (see SI). We thus remain focused on T = 258 K in the following determination of the adsorption isotherm.

Note also that using time-temperature equivalence, it is possible to extrapolate from the VFT laws the temperature evolution versus $\zeta$ at a fixed relaxation time, like $T_g$ at 100 s. This has been successfully applied to the DSC data in the SI building an equivalent calibration curve, $T_g(\zeta)$. Incidentally, if nanocomposite samples were found not to follow the VFT-law, this would imply a modification of the sample composition, e.g., a change in adsorption. This is not observed over the T-range covered in the present study, but one may expect that much higher temperatures would favor desorption due to the entropic gain of the small molecules.

Using eqs 1 and 4 for each data point in Figure 2, the partition of coating agent $m_{CA}$ into adsorbed mass ($m_{CA}^{ads}$) and mass freely distributed in the matrix ($m_{CA}^{free} = m_{CA} - m_{CA}^{ads}$) can be obtained. The adsorbed mass fraction then reads

$$m_{CA}^{ads} = m_{CA}\left(1 - \frac{\zeta_{free}}{\zeta}\right)(1 - \zeta_{free}) \tag{6}$$

The ratio $m_{CA}^{ads}/m_{CA}$ can be directly multiplied by the introduced surface density of silanes, 1.65 nm$^{-2}$ for a 12%w silane-to-silica ratio of C$_{18}$-silane, to obtain the effective coverage of the silica interface Θ in adsorbed molecules per nanometer squared. These results allow the construction of the adsorption isotherm shown in Figure 8, as a function of the concentration of free (i.e., non-adsorbed) silane in the matrix, $\zeta_{free}$.

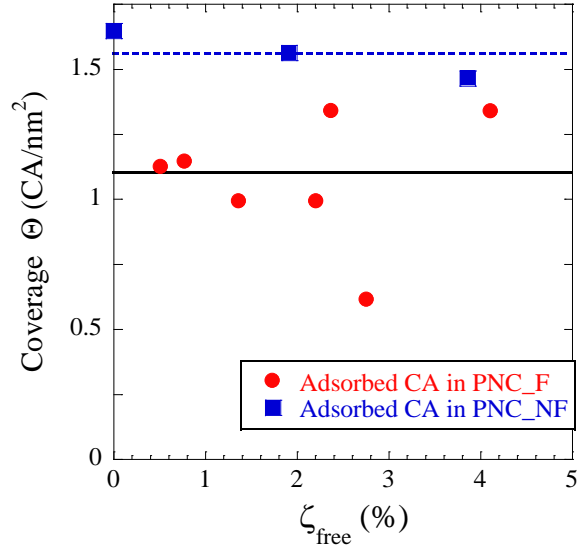

**Figure 8:** Adsorption isotherm of adsorbed amount Θ of coating agent per nm² vs. the concentration of free silane in the matrix polymer $\zeta_{free}$ in PNCs with various silica contents and non-grafted $C_{18}$-silanes in SB_F (squares) or SB_NF (circles) matrices. The solid black line represents the grafted density obtained from TGA when grafting in the NP suspension is activated. The blue dotted line is the average over the square points for PNCs in SB_NF.

Figure 8 represents the final summary of this work: via the plasticization effect of the silanes on the matrix, the amount of free silanes dispersed in the bulk has been determined, and by mass conservation, the remaining quantities on the silica interface could be deduced. This representation is exactly the one of an adsorption isotherm, where the adsorbed amount is reported as a function of the free silane concentration in the solvent, or here, in the polymer. The higher the free (matrix) concentration, the higher the pressure towards adsorption. Adsorption isotherms have been studied for almost a century, due to the obvious importance of interactions with surfaces and interfaces. [45-47] In the case of nanoparticles, adsorption is often studied for suspensions in solvents, as e.g. by Laurens et al who very recently proposed to correlate polymer adsorption with Hansen dispersibility parameters of silica in a large set of solvents. [48] Samples can then be centrifuged and supernatant concentrations determined, e.g., by optical spectroscopy, or surface tension measurements. [49-50] Typical examples are surfactant adsorption studies on silica NPs in water. [51-52] The shape of the isotherms gives crucial information on the local interactions, while the plateau provides the total adsorbed amount. Moreover, cooperative interactions between adsorbed species may further complicate the system, and thus the isotherm. [53-54] To our best knowledge, this is the first in-situ determination of the adsorbed quantity of silanes on the NP interfaces embedded in a polymer matrix. The inspection of the adsorption isotherm in Figure 8 shows that the plateau is already reached, although some scattering in the data is necessarily found. The plateau value for adsorption in functionalized polymer is ca. 1.1 nm$^{-2}$, with error bars due to the scattering of the points. This compares favorably with TGA, which evaluates the effective grafting density in case of activated grafting to also about 1.1 nm$^{-2}$. Thus, the comparison shows that even in the absence of grafting, a coating layer of density close to what would have been obtained with grafting is present with functionalized polymer.

Reaching the maximal coverage of the plateau-value in SB_F was to be expected, as the silane-to-silica ratio corresponds to what is typically used in industry to cover interfaces of NPs of this size. [15-16] Moreover, we have seen in Figure 7 that adding further silanes – this corresponds to multiplying the introduced silane surface density by the corresponding factor – does not further increase the adsorbed amount, but accumulates in the matrix. Besides some sticky patches necessary for understanding NP aggregation, there is thus no strong silane polycondensation on the nanoparticles. This a posteriori corroborates our assumption of the absence of the same phenomenon in the matrix. The corresponding points have been added onto the adsorption isotherm shown in Figure 8, and it is found that the interfacial coverage stays constant in presence of excess coating agents. If the matrix is replaced by the non-functionalized polymer, higher adsorption is observed and reported in Figure 8 (blue squares, average ca. 1.5 nm$^{-2}$) most probably due to the lower affinity of this polymer with the silanes. High enough excess silanes thus always plasticize the matrix, a phenomenon that must be kept in mind for formulation.

**Conclusion**

We have investigated the impact of silane coating agents on the segmental dynamics of the polymer chains in colloidal model silica nanocomposites by dielectric spectroscopy, and confirmed the observed effect independently by TM-DSC. Four different types of samples have been studied, with either grafted or adsorbed silanes ($C_8$, $C_{12}$, $C_{18}$), on NPs dispersed in functionalized (resp. non-functionalized)

SB. A BDS calibration curve of matrix plasticization on silica-free samples was established and used to show that only the total silane mass – and not its molecular weight – plays a role.

In presence of NPs, the dielectric response possesses several contributions, but the α-relaxation and thus its characteristic time $\tau_{max}$ can be isolated. In particular, while the shape of the relaxation time distribution function is slightly modified, the position $\tau_{max}$ is found to be robust for PNC-samples containing up to 20% of particles, as demonstrated by reference measurements with bare NPs. In other words, nanoparticles broaden the relaxation time spectrum, while plasticizers shift it. The calibration curve then provided access to the concentration of silanes in the matrix of a volume-fraction series of polymer nanocomposites, and by mass conservation to the mass adsorbed on the particle interface. Although the absence of matrix plasticization for silane molecules grafted on the silica NPs is trivial, such samples effectively demonstrate that BDS is a robust probe for silane partitioning. We have shown that due to the properties of the VFT-law, the measurements can be done at any temperature, and possibly rescaled to a calibration at a different temperature, as long as the samples do not change adsorption equilibria in the observed temperature range – such a behavior would show up through a deviation from the VFT-law. In parallel, the structure of the samples has been investigated by SAXS. Nanocomposites were found to be made of well-dispersed nanoparticles, with at most some small aggregates of a few NPs, for all surface modifications, i.e., silane grafting or adsorption. Aggregation was found to be slightly higher in case the grafting reaction was activated, presumably due to the simultaneous activation of polycondensation of silanes on the NP surface.

The main result of this article is that the plateau level of the adsorption isotherm within nanocomposites could be measured. The coverage of the NP surface is found to be very similar to grafted silanes, or even higher in the case of non-functionalized matrices which accept less silane. As experimental parameters of the silane-silica ratio have been fixed to industrial standards, it is reassuring to see that the plateau has indeed been reached, and that adding further silane does not increase the particle coverage. In industrial mixing units, most additives are introduced at once. Our work thus contributes to the understanding of the initial partitioning of the coating agents between matrix and the nanoparticle interfaces, before any possible surface modification reaction. It is hoped that our analysis will be useful for the quantification and optimization of chemical reactions at the interface in such complex samples, and may lead to reduced use of coating agents or catalysts, or more efficient macroscopic properties.

**ASSOCIATED CONTENT**

**Supporting Information.** The Supporting Information is available free of charge at https://XXX.
BDS analysis of PNCs with bare NPs, relaxation time distribution analysis of 10%-PNCs, temperature dependence of the calibration curve, and TM-DSC results.

**Acknowledgements.** The authors are thankful for support by the ANR NANODYN project, grant ANR-14-CE22-0001-01 of the French Agence Nationale de la Recherche. They acknowledge financial support from the European Commission under the Seventh Framework Program by means of the grant agreement for the Integrated Infrastructure Initiative N° 262348 European Soft Matter Infrastructure (ESMI). The SAXS experiments were performed on beamline ID02 at the European Synchrotron Radiation Facility (ESRF), Grenoble, France. Experimental support in BDS from Silvia Arrese-Igor (CSIC-UPV/EHU) is warmly acknowledged. The polymer was a gift from Synthos, and Radoslaw Kozak, Nathalia Meissner, and Pawel Weda (Synthos) are thanked for polymer synthesis.